\documentstyle[12pt]{article}
\begin{document}

\author{Guang-jiong Ni\thanks{%
E-mail: gjni@fudan.ac.cn} \\
Department of Physics, Fudan University \\
Shanghai, 200433, P.R.China}
\title{Where is the antimatter?\thanks{%
A lecture presented at the High Energy Physics Conference in China
Nationwide. (1998, 4.21-4.25 Chengde, Hebei, P.R.China).} }
\date{May 6, 1998 }
\maketitle

\begin{abstract}
The antimatter exists everywhere, but under suppressed state. So its
appearance is nothing but the various effects of special relativity.
\end{abstract}

The most remarkable features in high energy physics are:

(a) the equal existence of particle and antiparticle;

(b) the mutual transformation between various kinds of particles

For instance, the antiparticle of electron ($e^{-}$) is positron ($e^{+}$).
They have equal mass, opposite charge and are all stable. But once they
encounter each other, an ``annihilation'' process would occur and they
transform into a pair of photons:

\begin{equation}
e^{+}+e^{-}\rightarrow \gamma _1+\gamma _2
\end{equation}

The above process was first discovered by Chinese Physicist C. Y. Chao in
1930 (C. Y. Chao, Proc. Nat. Acad. of \hspace{0in} Sci. 16, 431(1930), Phys.
Rev. 36, 1519(1930)). He measured that the energy of one photon equals
approximately to the rest energy of electron: $E_{\gamma _1}\simeq E_{\gamma
_2}\simeq m_ec^2=0.511MeV$. Two years later, Anderson verified definitely
the existence of positron (C. D. Anderson, Science 76,238(1932)). See Ref
[1].

The research in Astrophysics reveals that the quantity of antimatter
in the universe is far less than that of matter with a
ratio $< 10^{-9}$ approximately. The tiny asymmetry may have
important meaning in the combination of particle physics with cosmology. The
enormous energy source of QUASAR, the remote celestial body discovered since
1960, may have some relevance to the annihilation of matter and antimatter.
Besides, the enormous energy release in the supernova explosion might also
be relevant to antimatter according to some theoretical conjecture.. But to
our knowledge, no definite answer is achieved.

The aim of this paper is to propose an alternative point of view. The
antimatter is existing everywhere, including everyone's body. Though its
ratio to matter is extremely small, its absolute quantity is rather large.
The reason why we don't feel it is as follows. The antimatter is under a
suppressed state, so instead of opposite charge, its appearance is precisely
the special relativity (SR) effect familiar to every physicist.

Since the discovery of parity violation in 1956 and the CP violation (i.e.,
the nonconservation of T inversion whereas CPT theorem remains valid) in
1964, physicists have been gradually accepting the definition relating the
particle state $|a>$ to its antiparticle state $|\overline{a}>$ as a
combined CPT transformation rather than C transformation:

\begin{equation}
|\overline{a}>=CPT|a>
\end{equation}

Performing the inversion of C, P and T independently on the wave function $%
(\psi _{e^{-}})$ of an electron, one sees that the combined effect of CPT
inversion is essentially ascribed to the transformation $\overrightarrow{x}%
\rightarrow -\overrightarrow{x},t\rightarrow -t$. (There is a complex
conjugate transformation in C inversion as well as in T inversion, they
cancel each other). Thus

\begin{equation}
\psi _{e^{-}} \simeq \exp \{\frac{i}{\hbar}
(\vec{p}\cdot \vec{x}-Et)\}
\end{equation}
turns to the wave function of a positron
\begin{equation}
\psi _{e^{-}} \simeq \exp \{\frac{i}{\hbar}
(\vec{p}\cdot \vec{x}-Et)\}
\end{equation}
automatically, where $\overrightarrow{p}$ is the momentum of $e^{-}$ or $%
e^{+}$ and $E$ $(>0)$ is the energy individually. Eqs (3) and (4) denote
that the difference between wave functions of particle and antiparticle is
ascribed to the opposite sign in their phases. This point of view, to our
knowledge, was proposed by Schwinger et al [3] in quite early time. However,
it was overlooked by the majority of physics community.

It seems to us that once we go one step further from the above point of
view, an interesting and important thing will occur. Being some special kind
of wave function, Eqs (3) and (4) show a symmetry:

\begin{equation}
\Psi _{e^{-}}(\overrightarrow{x},t)\rightarrow \Psi _{e^{-}}(-%
\overrightarrow{x},-t)=\Psi _{e^{+}}(\overrightarrow{x},t)
\end{equation}

Now it is the time to go forward ``from the special case to the general
case''. Suppose that the symmetry shows by Eq (5) is a general law of
nature, i.e., any matter comprises two opposite degrees of freedom,
described by wave functions $\theta (-\overrightarrow{x},-t)=\chi (%
\overrightarrow{x},t)$ with the symmetry

\begin{equation}
\theta (-\overrightarrow{x},-t)=\chi (\overrightarrow{x},t)
\end{equation}

Correspondingly, the coupling equation of $\theta $ and $\chi $ should be
invariant under the transformation $(\overrightarrow{x}\rightarrow -%
\overrightarrow{x},t\rightarrow -t)$ and the substitution Eq.(6).

Based on the above postulate, one is able to derive, in an inertial
coordinate system without resort to the Lorentz transformation, the
Klein-Gordon equation and Dirac equation ([4], [5]). A new stationary
Schr\"{o}dinger equation for many particle system with relativistic
modification is also derived ([6]). Hence we see that the wave function of
an electron shown by Eq.(3) is actually a coherent superposition of two
ingredients, $\theta $ and $\chi $. In accompanying with the enhancement of
momentum $\overrightarrow{p}$, $\chi $ increases from zero gradually till a
limit is \hspace{0in}achieved at $\overrightarrow{p}\rightarrow \infty $, (
the velocity $v\rightarrow c$ ):

\begin{equation}
\lim_{v \rightarrow c} \left| \chi \right| =\left| \theta \right|
\end{equation}

The subtlety lies in the fact that there is \hspace{0in}always an unequality 
$\left| \theta \right| >\left| \chi \right| $ in $\Psi _{e^{-}}$ and they
don't exhibit the symmetry (6) explicitly. The explanation is as follows.
According to its instinct, $\theta $ demands the space-time evolution
direction in its phase as shown in Eq.(3) whereas $\chi $ demands the
opposite one as shown by Eq.(4). However, being the ``slave'' in $\Psi
_{e^{-}}$, $\chi $ has to obey the disposition of ``master'' $\theta $ , so
both of them display the phase evolution law as shown by Eq.(3). Of course, $%
\chi $ is not willing to do like this. But what he can do is to try to hold
the electron back. Hence the inertial mass of electron increases without a
limit. Meanwhile, the reading of ``moving clock'' in accompanying with $%
\theta $ ( or $\chi $ ), according to Eq.(3) ( or (4) ) in instinct, is read
out ``clockwise'' ( or ``anticlockwise'' ). Now $\theta $ dominates the
electron, so its clock reading is still clockwise. However, it becomes
slower and slower with the enhancement of ingredient $\chi $ [7].

Performing a space-time inversion ($\overrightarrow{x}\rightarrow -%
\overrightarrow{x},t\rightarrow -t$ according to our new definition) on the
electron wave function, one finds that $\theta \rightarrow \chi _c$, $\chi
\rightarrow \theta _c$ with $\left| \chi _c\right| >\left| \theta _c\right| $
, ( the subscript $c$ refers to antiparticle state ), bringing Eq.(3) to
Eq.(4), i.e., an electron changes to a positron.

Therefore, a particle is always not pure. When a particle is set into motion
from the rest state, the antiparticle ingredient is excited coherently
inside it. The various effects in special relativity are nothing but the
various reflections of antiparticle ingredient which is under a subordinate
status and is just displaying its presence tenaciously. One of its
reflections is as follows. As is well known, In nonrelativistic quantum
mechanics (QM), the modulus square of wave function can be explained as
``the probability density of particle at position $\overrightarrow{x}$ when
the measurement is made'' by Born statistical interpretation. However, it
was still hard to understand why a wave packet of freely moving particle
would spread unceasingly? In Ref [8] we try to interpret the wave function
as some abstract representation of ``contradiction field'' between particle
and its environment, with $\overrightarrow{x}$ being the flowing coordinate
of ``field'' rather than the position coordinate of ``point particle''. Then
as one assumes first the external potential field $V(x)=0$ ( no interaction
between particle and environment ), next one supposes that $\psi (x,0)$ has
a wave packet distribution at $t=0$ , so consequently, the wave packet has
no way but to spread and approach to zero. The situation becomes quite
different for Klein-Gordon equation or Dirac equation, where the wave packet
does not spread at $v\rightarrow c$ , rather, it suffers from a boosting in
accompanying with the Lorentz contraction. In our point of view, it is just
a reflection of coexistence of two field, $\theta $ and $\chi $ , the latter
being a little bit smaller. They entangle together to ensure the inner
stability of a particle, which behaves like a classical particle again at $%
v\rightarrow c$. This is the reason why we can devise the pulse working mode
for a high energy particle accelerator.

Einstein established the SR by two papers in 1905. In the first paper he
raised two ``relativistic postulates'', the ``postulate of relativity'' and
the ''postulate of constancy of speed of light''. Then in the second paper
he derived the mass-energy relation $E^2=mc^2$ ingeniously by means of the
Lorentz transformation in the propagation of light. Thus the ``light''
occupies a unique position in SR. From our point of view, a photon with
right circular polarization could be viewed as a ``particle'' while the left
circular polarization one is a ``antiparticle''. So a linear polarization
photon could be viewed as an superposition of particle and antiparticle
states each with 50\% ingredient. A photon is always an extremely
relativistic particle. In some sense, when Einstein established SR by
``light'', he was also making use of the method ``from the special case to
the general case''.

The great success of SR and QM makes physicists familiar with the deduction
method which emphasizes the way from the general case to special case.
However, the other way from the special case to general case, i.e., the
analysis and induction method is also very important and should not be
neglected. Only the combination of these two methods will be beneficial to
the further development of physics.

\vspace{0in}\hspace{1.1in}

\hspace{1.1in}References

[1] Bing-an Li and C. N. Yang, Int. Jour. Mod. Phys A, 4, No.17, (1989) 4325.

[2] T. D. Lee and C. S. Wu, Annual Rev. Nucl. Sci. 15 (1965) 381.

[3] J. Schwinger, Proc. Nat. Acad. Sc. U.S. 44 (1958) 223;

\hspace{0in} \hspace{0in} \hspace{0in} \hspace{0in} \hspace{0in}E.J.
Konopinski and H.M. Mahmaud, Phys. Rev 92 (1953) 1045.

[4] G-j Ni, The relation between space-time inversion and
particle-antiparticle transformation, Journal of Fudan University (Natural
Science), 1974, No.3/4,125.

[5] G-j Ni and S-q Chen, On the essence of special relativity, ibid, 35
(1996) 325; Internet, hep-th/9508069 (1995).

[6] G-j Ni and S-q Chen, Relativistic Stationary Schr\"{o}dinger equation
for many particle system, ibid, 36 (1997) 247; G-j Ni, hep-th/9708156 (1997).

[7] G-j Ni, To enjoy the morning flower in the evening------Is special
relativity a classical theory?

\hspace{0in} \hspace{0in} \hspace{0in} \hspace{0in} Kexue ( Science), 50
(1998), No. 1, 29; quant-ph/9803034.

[8] G-j Ni, \hspace{0in}To enjoy the morning flower in the
evening------where is the subtlety of quantum mechanics? ibid, No. 2, 38;
quant-ph/9804013.

\end{document}